%% file: main.tex
\documentclass[letterpaper, 10 pt, conference]{ieeeconf}  

\IEEEoverridecommandlockouts                              

\let\labelindent\relax
\usepackage{amssymb}
\usepackage{amsmath}
\usepackage{enumitem}
\usepackage{graphicx,subcaption}
\usepackage{times} 
\usepackage{algorithm, algorithmic}
\usepackage{soul}
\usepackage{color}
\usepackage{svg}
\usepackage{dblfloatfix}
\usepackage[noadjust]{cite}

\title{\LARGE \bf
Task Allocation with Load Management in Multi-Agent Teams
}

\author{Haochen Wu$^{1}$, Amin Ghadami$^{1}$, Alparslan Emrah Bayrak$^{2}$, Jonathon M. Smereka$^{3}$, and Bogdan I. Epureanu$^{1}$%
\thanks{This work was supported through the Automotive Research Center, University of Michigan, Ann Arbor, MI, USA. Distribution A: Approved for public release; distribution unlimited. OPSEC\# 5781.}
\thanks{$^{1}$Haochen Wu, Amin Ghadami, and Bogdan I. Epureanu are with the University of Michigan, 
        Ann Arbor, MI 48109, USA
        {\tt\small \{haochenw, aghadami, epureanu\}@umich.edu}}%
\thanks{$^{2}$Alparslan Emrah Bayrak is with the School of Systems and Enterprises, Stevens Institute of Technology,
        Hoboken, NJ 07030
        {\tt\small ebayrak@stevens.edu}}%
\thanks{$^{3}$Jonathon M. Smereka is with the US
Army CCDC Ground Vehicle Systems Center,
        Warren, MI 48397, USA
        {\tt\small jonathon.m.smereka.civ@army.mil}}%
}

\begin{document}

\maketitle
\thispagestyle{empty}
\pagestyle{empty}

\input{sections/Abstract}

\section{Introduction}
\input{sections/Introduction}

\section{Methods}
\label{sec:MET}
\input{sections/Methods.tex}

\section{Results}
\label{sec:RES}
\input{sections/Results.tex}

\section{Conclusions}
\input{sections/Conclusions.tex}

\addtolength{\textheight}{-12cm}

\bibliographystyle{IEEEtran}
\bibliography{ref.bib}

\end{document}

%% file: sections/Abstract.tex
\begin{abstract}
In operations of multi-agent teams ranging from homogeneous robot swarms to heterogeneous human-autonomy teams, unexpected events might occur. While efficiency of operation for multi-agent task allocation problems is the primary objective, it is essential that the decision-making framework is intelligent enough to manage unexpected task load with limited resources. Otherwise, operation effectiveness would drastically plummet with overloaded agents facing unforeseen risks. In this work, we present a decision-making framework for multi-agent teams to learn task allocation with the consideration of load management through decentralized reinforcement learning, where idling is encouraged and unnecessary resource usage is avoided. We illustrate the effect of load management on team performance and explore agent behaviors in example scenarios. Furthermore, a measure of agent importance in collaboration is developed to infer team resilience when facing handling potential overload situations.

\textit{Keywords}: Multi-Robot Systems, Task Planning, Reinforcement Learning, AI-Based Methods, Cooperating Robots
\end{abstract}

%% file: sections/Introduction.tex
Multi-agent systems have significant potential in generating management-level strategies for real-world large-scale operations such as humanitarian relief, search and rescue, transportation, warehouse management, and various other scenarios that involve teaming of heterogeneous agents or human-autonomy collaboration. In a team of collaborative agents, instead of being managed by a centralized system, decentralized agents have equal authorities in task planning and execution and ensuring operation effectiveness, with better team resilience in unexpected situations and scalability. With the help of emerging computing systems, recent advances in multi-agent task allocation \cite{MATAreview,MATA2,MATA3} have deliberated on distributed systems and deep reinforcement learning methods \cite{DEC-POMDP,HOOSHANGI2017,DQN,omidshafiei2017deep,HT-DEC-POMDP} for complex scenarios in addition to centralized approaches \cite{Nair2013,emam2020adaptive,QMDPNet}. Most coordination and collaboration strategies in efficient task allocation \cite{MATARL} define the objective as accomplishing tasks \cite{HT-DEC-POMDP} or reaching task locations \cite{omidshafiei2017deep, QMDPNet} in fewer steps. Although these approaches could sufficiently accomplish the problem objective, there is a lack of consideration of risk management since unanticipated situations might require additional strategic amendments in real-world operations. Agents in a team should be intelligent enough to prepare for the unknown.

To mitigate such risks and potentially allow agents to be available for unexpected events, some agents should idle. Being constantly active would overload the agents, would be especially harmful to human agents, and would cost extra resources. In the literature, load is handled as management or load allocation problems \cite{ResourceManage,ResourceAllocation,ResAlloMA} in the field of multi-agent computing. In human-robot collaboration, load usually refers to the workload measured by human heart-rate variability \cite{MAWorkload}, serving as a feedback to reduce stress and improve overall team performance \cite{AdaptiveWLA}. In the context of multi-agent task allocation in dynamic environments, the idling time is a performance measure to indicate task load balance \cite{AuctionTAforMR}, a precondition/constraint to perform other tasks \cite{Noureddine2017MultiagentDR}, or a time indicator for inoccupation \cite{idleinoccu}.

In addition, optimization-based methods for load reduction have been studied as centralized strategies in the field of production scheduling \cite{prodSche}, energy management \cite{energyManage}, and cloud computing \cite{cloudComp}.
Limited research has explored the effect of idling as an option and to investigate the effects of load reduction on performance for decentralized multi-agent teams. Therefore, this work considers idling behaviors as an incentive to reduce task load and save resources represented by agent capabilities and task-reassignment.

In this paper, we leverage our previous work \cite{HT-DEC-POMDP} on task allocation for heterogeneous multi-agent teams to model task load management based on agent preference in decision-making as a secondary objective along with maximizing team performance.
Multi-agent teams with a reduced load respond more reliably in emergencies that require extra resources, or in situations where one of the team members is compromised \cite{resilientTA}. An optimal idling strategy for agents may have negligible impact on team effectiveness, but makes a team more resilient in unexpected situations. We identify idling strategies by quantifying the importance of each agent in a team.
The proposed work also accounts for individual agent preferences in task assignment. The analysis on the impact of load reduction and the importance of collaboration highlights the benefits of the proposed decision-making framework.

The contributions of this work are:
\begin{itemize}
    \item Efficient learning of load reduction in multi-agent teams with heterogeneous capabilities without compromising team performance,
    \item Embedding agent preferences in decision-making by customizing individual reward functions,
    \item Inferring team resilience by characterizing the importance of each agent on team performance.
\end{itemize}

%% file: sections/Methods.tex
To enable intelligent teaming of heterogeneous agents in task allocation problems, it is assumed that all agents possess individual decision-making processes instead of following commands. In lieu of being subordinate members, agents are equally responsible for strategic planning and task execution. We characterize the heterogeneity in agent decision models by allocation preferences which can be customized by agent reward functions. While agents have the same type of attributes to handle tasks, sense the environment, and communicate information, the capability levels for each attribute are heterogeneous. Such heterogeneous and specialized agents need to collaborate to complete all the tasks within fewer steps. An efficient learning method is desired to achieve intelligent teaming behaviors with load management, by which agents stay idle when they are not needed without compromising the overall team effectiveness.

\subsection{Problem Formulation}
\label{sec:prob_form}
The decision-making process for a team of heterogeneous agents with load management inherits the formulation of our previous work on heterogeneous teaming decentralized partially observable markov decision process (HT Dec-POMDP) \cite{HT-DEC-POMDP} and incorporate two additional features allowing intelligent load management: 1) agent's option to idle, and 2) penalty on unnecessary task assignment depending on capabilities and current task assignment. The proposed framework heterogeneous teaming with load management (HTLM Dec-POMDP) is defined as a tuple $(\bar{G},\alpha,S,\bar{A},\mathcal{T},\bar{R},Z,\mathcal{O},M,\mathcal{I},h,b^0,\gamma)$ where:

\begin{itemize}
    \item $\bar{G}:=\{g_1,\dots,g_p,\bar{g}_1,\dots,\bar{g}_q\}$ is a finite set of $p$ tasks and $q$ idle locations
    \item $\alpha:=\{\alpha_1,\dots,\alpha_n\}$ is a finite set of $n$ agents
    \item $S:=S^D\times S^C, s\in S,$ is the overall state that is factored into $p$ tasks and described by task demand/severity levels $S^D:=S^D_1\times\dots\times S^D_p$ and joint agent capabilities $S^C:=S^C_1\times\dots\times S^C_p$
    \item $\bar{A}:=\times_i \bar{G}, a\in \bar{A},$ is the set of joint assignment decision by assigning $\alpha_i$ to $g_j$
    \item $\mathcal{T}:=Pr(s'|s,a)$ is the state transition probability
    \item $\bar{R}:=f_R(a_o,b(s),a)$ is the reward function, where $a_o$ records the previous assignment, $b$ is the belief defined as the probabilistic distribution over the problem state space, $b'$ is the belief for the next state
    \item $Z:=\times_i Z_i, z\in Z,$ is the set of joint observations
    \item $\mathcal{O}:=Pr(z|s',a)$ is the observation probability
    \item $M:=\times_iM_i, m\in M,$ is the set of joint information, where $M_i$ represents the state information collected by agent $\alpha_i$
    \item $\mathcal{I}=Pr(m|s',a,z)$ is the information probability
    \item $h$ is the planning time horizon
    \item $b^0$ is the initial belief 
    \item $\gamma$ is the discount factor
\end{itemize}

This formulation follows the fundamental modeling of environment dynamics in HT Dec-POMDP \cite{HT-DEC-POMDP}, and explicitly defines various types of attributes (task-specific, sensing, and communication capabilities) for agents and the corresponding probabilities that describe the stochastic environment and the agent's perception of the environment and information. In many studies, agents stay active throughout the operations, which might lead to excessive usage and risk exposure. Instead, HTLM Dec-POMDP provides agents with the option to stay idle by including idle locations as part of the assignment decisions described by $\bar{G}$ and $\bar{A}$. 
Note that an action of an agent refers to the task assignment, i.e. the agent executes the assigned task. In addition, the reward function $\bar{R}$ not only defines the rewards for completing the tasks, but it also penalizes unnecessary assignment and reassignment to induce agent idle behaviors without compromising team performance. Since the formulation is decentralized, agents have their own reward functions representing their understandings of the situation and preferences on the task assignment, which might be heterogeneous among the team members.

The goal of the formulation is, for each agent, to find an optimal series of task allocation actions or policy $\pi_*$ that maximizes the operation performance so that $\pi_*=argmax\{V(\pi)\}$. The performance is represented as the policy value $V(\pi)$ defined as the expected total future discounted reward.
\begin{equation}
    V_\pi=\mathbb{E}[\sum_{t=0}^{t=h} \gamma^t f_R(a^{t-1},b^t(s),a^t)|\pi,b^0].
\end{equation}
To give more incentives for agents to accomplish tasks at earlier time steps, the reward is discounted by time ($\gamma^t$). At the beginning of an operation, each agent has an initial belief $b^0(s)$ over the operation state. At each time step of a time horizon $h$, agents execute decisions $\bar{A}$, receive partial observations $Z$, and receive information from others ${M}$. After experiencing the action-observation-information histories, agents update their beliefs over the state following Eq. (\ref{eq:belief_update}) with capability-dependent transition probability $\mathcal{T}$, observation probability $\mathcal{O}$, and information accuracy $\mathcal{I}$ \cite{HT-DEC-POMDP} in order to make the decisions for the next time step 
(the prime symbol denotes the next time step), namely
\begin{align}
    \label{eq:belief_update}
   b'(s')&= Pr(s'|m,z,a,b)\nonumber\\
   &\propto\mathcal{I}(s',a,z,m)\mathcal{O}(s',a,z)\sum_s \mathcal{T}(s,a,s')b(s).
\end{align}
The approaches of defining the reward functions and learning the optimal policies based on beliefs are discussed in Sections \ref{sec:preference} and \ref{sec:learning}.

\subsection{Embedding Agent Preference in Reward Shaping}
\label{sec:preference}
Defining a reward function, also referred to as reward shaping, plays a critical role in reinforcement learning. A carefully designed reward function can expedite the training process while satisfying the problem objectives. In our decentralized task allocation problem, there are two objectives: 1) completion of tasks and 2) load management. Moreover, reward functions could be customized to model agent preferences, meaning that agents can have different reward functions depending on their willingness to take risks and to reassign tasks. Providing rewards based on the actual operation state \cite{Noureddine2017MultiagentDR} or system score \cite{DQN} is not applicable since there does not exist a centralized informer. Therefore, agents reward themselves based on their belief over the status of the demands and the environment.

\begin{equation}
    \label{eq:htlm_reward}
    f_R(a_0, b(s), a)=R_{comp}+R_{LM},
\end{equation}
\begin{equation}
    \label{eq:comp_reward}
    R_{comp}=W^T_lR_l,
\end{equation}
\begin{equation}
    \label{eq:LM_reward}
    R_{LM}=R_{II}(b(s))+T_{RP}(a_0,a).
\end{equation}

Eq. (\ref{eq:htlm_reward}) defines the reward function for each agent in the proposed HTLM Dec-POMDP.
The agent preference in this context refers to the customizable reward shaping for each agent. In particular, different agents might have different confidence in mission completion, which is modeled in $R_{comp}$; $R_{II}$ models agents' different incentives on idling; and $T_{RP}$ is utilized to describe agent preferences on certain tasks or the locations of the tasks.
$R_{comp}$ is a belief-dependent reward computed based on the belief state and the belief reward threshold $b_{th}$. By deciding how much and how generous each agent informs the reward of mission completion based on partial completion weight $w$ and belief reward threshold $b_{th}$ respectively, $W_l=f(b(s);w,b_{th})$ and $R_l$ provides the reward for each task severity level $l$ \cite{HT-DEC-POMDP}.
The function $f(b(s);w,b_{th})$ is defined such that agents receive partial reward for decreasing the demand levels to level $l$ with maximum reward received when all the tasks are completed, i.e. when $l=0$. For example, if an agent believes, with a probability of at least $b_{th}$, that the severity levels of half of the tasks ($w=0.5$) is at level 2 $(l=2)$ and the other half of the tasks are completed $l=0$, then the $R_{comp}$ reward is computed as $R_{comp}=0.5R_0+0.5R_2$. The rewards $R_l$ depend on the studied scenario and need to be tuned for an efficient training, but $R_j>R_i,  j<i$ should be used to guide agents towards completing tasks.


With only $R_{comp}$ reward, the team of agents learn to accomplish the desired tasks effectively, but there exists a number of optimal solutions. The deep learning approach randomly converges to one of the optimal solutions, but fails to avoid task overload. The second part of the reward function is the reward of load management $R_{LM}$ in HTLM Dec-POMDP defined in Eq. (\ref{eq:LM_reward}). $R_{LM}$ has two features - idle incentive $R_{II}$ and reassignment penalty $T_{RP}$ - in order to achieve the following behaviors without the compromise of operation effectiveness: idling as much as possible while avoiding excessive capability usage and unnecessary task reassignment. Load management serves as the secondary objective after mission completion. Therefore, the maximum possible $R_{LM}$ is much smaller than $R_{comp}$. Additionally, 
the belief-dependent $R_{II}$ is positive (reward) if the agent is assigned to idle at one of the locations and negative (penalty) if the capability level of the agent $i$ is much higher than the believed severity level of the assigned task $j$ ($c_i>>b(s_j)$), penalizing the assignment of an agent with excessive capability to an easy task.
The reassignment penalty $T_{RP}$ is a cost matrix defining the cost of changing task assignment from $a_0$ to $a$. 
The task reassignment penalty becomes the traversing cost when only the locations of tasks are used to calculate the penalty, i.e. when there is no cost of switching tasks at the same location.

In the task set $\bar{G}$, there are idle locations corresponding to the task locations. We assume that there is no cost of going into and out-of idling within the same location, and there is penalty for being reassigned to a different location. A sample $T_{RP}$ cost matrix for 2 locations is shown in Table \ref{tb:cost_matrix}, where the traversing cost between two locations is $tc\leq0$.
\begin{table}[ht]
\caption{Two-Location Traversing Cost Matrix}
\centering
    \begin{tabular}{|c|c|c|c|c|}
         \hline
         & Loc 1 & Loc 2 & Loc 1 & Loc 2 \\
         & Task & Task & Idle & Idle\\
         \hline
         Loc 1 Task & 0 & $tc$ & 0 & $-\infty$\\
         Loc 2 Task & $tc$ & 0 & $-\infty$ & 0\\
         Loc 1 Idle & 0 & $tc$ & 0 & $-\infty$\\
         Loc 2 Idle & $tc$ & 0 & $-\infty$ & 0\\
         \hline
    \end{tabular}
\label{tb:cost_matrix}
\end{table}

\noindent If only one idling action is considered for multiple locations with no cost of going into idling and out-of idling, agents can bypass the reassignment penalty by going idle and being assigned to a different task with two steps, which has a indirect impact on team effectiveness. Including multiple idle locations could explicitly define the reassignment penalty and have a direct impact on agent's decision for each time step. To avoid the additional computation due to the larger action space, we generate a reduced action space for each agent $\hat{A}=\{g_1\dots g_p, \bar{g}_o|a_o\}$ to restrict idling to a different location. Only the idle location $\bar{g}_0$ that has the same location as the previous assignment $a_0$ is included in the reduced action space while other idle locations are eliminated. For example, as shown in Table \ref{tb:cost_matrix}, the actions with $\infty$ cost would be eliminated given the previous assignment. The reduced action space for each agent at each step is $p+1$, where $p$ is the number of tasks and the additional action is idling at the same location. Note that by removing idle locations and setting $R_{LM}=0$, the framework would reduce to HT Dec-POMDP. The investigation on the effect of $R_{LM}$ on team behaviors and performance when idle actions are available is discussed in Section \ref{sec:R_LM_effect}.

\subsection{Decentralized Deep Q-Learning for Load Management}
\label{sec:learning}
To achieve load management by avoiding excessive capability usage and unnecessary assignment, an agent has to acknowledge the severity of the operation and record the previous assignment. To approach the optimal solution, the proposed method extends the decentralized deep Q-learning with beliefs \cite{HT-DEC-POMDP} as shown in Algorithm~\ref{alg:decdql_LM} (underlined parts indicate the contribution of the proposed learning method for load management), and an agent makes intelligent decisions based on a knowledge state $\mathcal{K}=\{b(s),a_o\}$ which consists of its belief of the operation state $b(s)$ and the previous assignment $a_o$. The knowledge state update (Alg.~\ref{alg:decdql_LM} Line 8) inherits the Markov property since the next belief $b'(s')$ only depends on the current belief $b(s)$ as described in Eq. (\ref{eq:belief_update}), and the decision made becomes the previous assignment in the next time step.

\begin{algorithm}
\caption{Dec Deep Q-Learning for Load Management}
\begin{algorithmic}[1]
\label{alg:decdql_LM}
\STATE Initialize replay memories $D$ and Q-Networks with random weights \textit{for all agents}
\FOR{each training episode}
\STATE Reset operation and \underline{knowledge state $\mathcal{K}$} \textit{for all agents}
\FOR{each operation step}
\STATE \textit{For all agents}
\STATE Select decisions \underline{from $\hat{A}$} with $\epsilon$-greedy and execute
\STATE Observe and communicate observations
\STATE \underline{Update $\mathcal{K}=\{b(s),a_o\}$}
\STATE \underline{Compute $f_R(\mathcal{K},a)$ and store $\mathcal{X}$ into $D$}
\STATE Train Q-Network with a randomly sampled minibatch of transitions in $D$
\ENDFOR
\ENDFOR
\end{algorithmic}
\end{algorithm}

Given the reward function discussed in Section \ref{sec:preference}, the discount factor $\gamma$, and the time horizon $h$, 
the action-value or Q-Value $Q^{\pi}(\mathcal{K},a)$ given a policy is defined as the maximum possible expected total reward when having knowledge state $\mathcal{K}^t=\mathcal{K}=\{b, a_o\}$ and taking action $a=\pi(\mathcal{K})$ under policy $\pi$.
\begin{equation}
    \label{eq:Q*ba}
    Q^{\pi}(\mathcal{K},a)=\underset{\pi}{\max}\mathbb{E}[\sum_{t=t'}^{t=h}\gamma^{t-t'}f_R(\mathcal{K}^t,a^t)|\mathcal{K}^t=\mathcal{K}, a^t=\pi(\mathcal{K})].
\end{equation}

\noindent Under \textit{Bellman Equation}, the Q-Value $Q^{\pi}(\mathcal{K},a)$ can be rewritten using the next knowledge-action pair $(\mathcal{K}',a')$.
\begin{equation}
    \label{eq:bellman}
    Q^{\pi}(\mathcal{K},a)=\mathbb{E}_{\mathcal{K}'}[f_R(\mathcal{K},a)+\gamma \text{max}Q^{\pi}(\mathcal{K}',a')|\mathcal{K},a].
\end{equation}

\begin{figure}[!t]
    \centering
    \vspace{2mm}
    \includegraphics[width=0.48\textwidth]{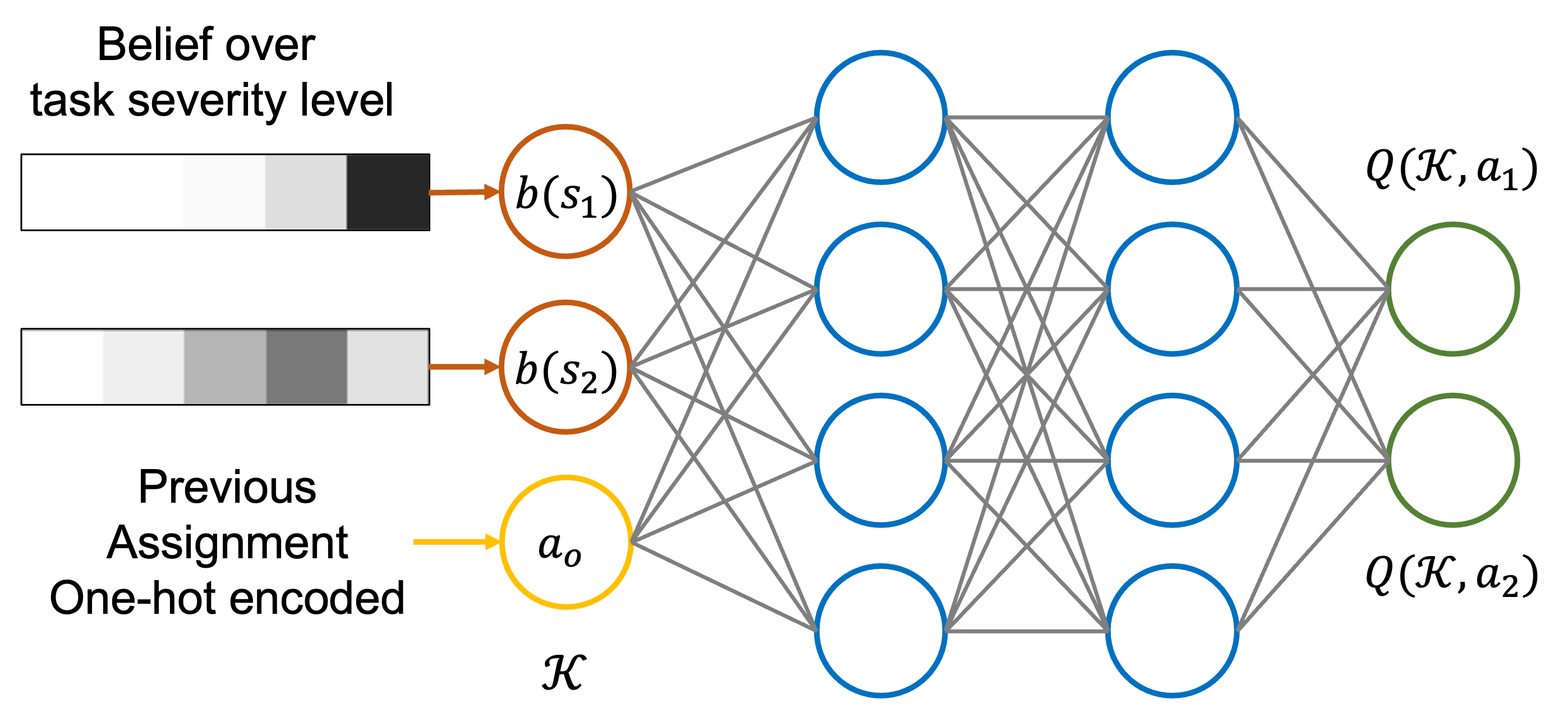}
    \caption{Schematic of a Q-Network with an agent's belief over severity levels of 2 tasks and previous assignment as input and Q-Values of 2 actions as output. In the proposed approach, agent's knowledge state $\mathcal{K}$ is used as an input to the Q-Network and the outputs are the corresponding Q-Values $Q(\mathcal{K},\cdot)$ in the action space.}
    \label{fig:q_network_K}
\end{figure}

Instead of exploring the full continuous knowledge state space, a neural network as shown in Fig. \ref{fig:q_network_K}. The Q-Network $Q(\mathcal{K},a;\theta)$ is used to approximate the Q-Value with the best policy $Q(\mathcal{K},a;\theta)\approx Q^{\pi}(\mathcal{K},a)$. The optimal policy for each agent can then be computed by $\pi^*(\mathcal{K})=\text{argmax}_{\hat{a}} Q(\mathcal{K},a;\theta)$. An agent selects the action with the highest Q-Value from the reduced action space $\hat{A}$. Recall that $\hat{A}\subseteq \bar{A}$. The input of the Q-Network is the knowledge state which includes the belief state over all task severity levels in probability space and the one-hot encoded previous assignment. The total number of input neurons is $pL+|\bar{G}|$, with $p$ being the number of tasks and $L$ being the number of task severity levels. The output of the Q-Network is the Q-Value for each action, and the size of the output is $|\bar{G}|$. The Q-Network output includes all tasks and idle locations since agents can be assigned to any of them throughout an operation. The reduced action space is helpful when generating random actions and selecting the optimal action (Alg.~\ref{alg:decdql_LM}, Line 6) so that agents do not waste effort on reassigning to the location with $\infty$ cost. The benefit of using a deep learning approach in decentralized multi-agent teams is that the state space and the actions space are linear with respect to the number of tasks.

During each training iteration $i$, a set of transition data $\mathcal{X}=<\mathcal{K},a,\mathcal{K}',f_R(\mathcal{K},a)>$ is collected into replay memories (Alg.~\ref{alg:decdql_LM}, Line 9). With the Bellman Eq. (\ref{eq:bellman}), the Q-Network $Q(\mathcal{K},a;\theta)$ should converge to a target Q-Value $y_i=f_R(\mathcal{K},a)+\gamma \text{max}Q(\mathcal{K}',a';\hat{\theta}_i)$, where the parameter of the target Q-Network $\hat{\theta_i}$ is updated with $\theta$ less frequently. The difference in Q-Value and the target Q-Value is the loss $\mathcal{L}_i$ to be minimized to reach Bellman Optimality as defined in:
\begin{equation}
    \label{eq:loss}
    \mathcal{L}_i(\theta_i)=\mathbb{E}_\mathcal{X}[y_i(\mathcal{K}',a',\hat{\theta}_i)-Q(\mathcal{K},a;\theta_i)].
\end{equation}

%% file: sections/Results.tex
\begin{figure*}[!t]
    \centering
    \begin{subfigure}{0.9\textwidth}
        \centering
        \includegraphics[width=0.9\textwidth]{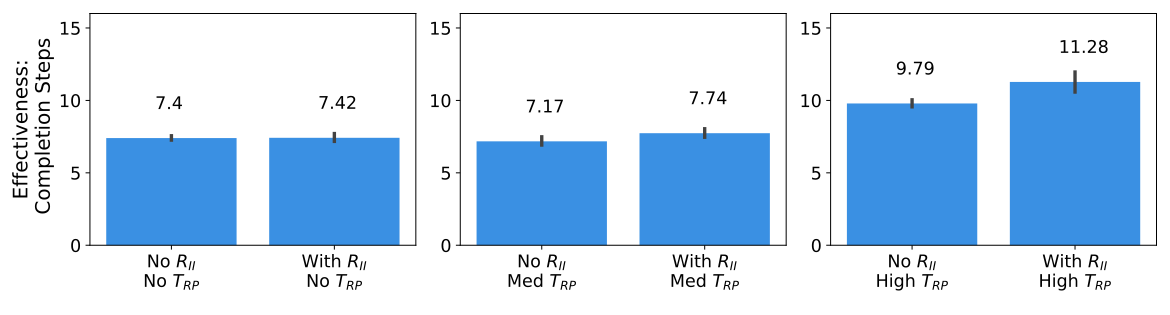}
        \vspace{-3mm}
    \end{subfigure}
    \vspace{-3mm}
    \begin{subfigure}{0.9\textwidth}
        \centering
        \includegraphics[width=0.9\textwidth]{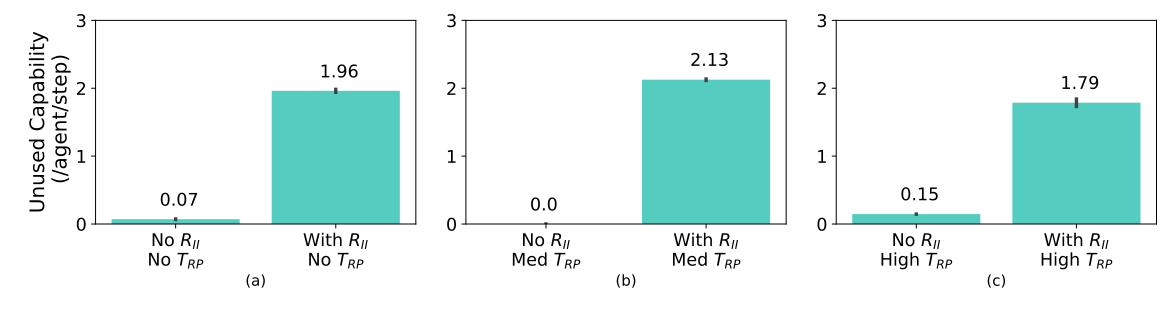}
    \end{subfigure}
    \caption{Effect of idle incentive $R_{II}$ on operation effectiveness and capability usage with (a) no reassignment (b) medium reassignment and (c) high reassignment penalty $T_{RP}$. The performance evaluation includes 500 realizations, and each bar plot shows the mean and the standard error.}
    \label{fig:idle_effect}
\end{figure*}

We analyze the teaming behaviors of the proposed HTLM Dec-POMDP framework in a disaster relief operation scenario. The operation has a team of 5 agents and 2 disaster sites. At each site, there are two different but conditionally dependent tasks: extinguishing fire and rescuing people from the burning site (4 tasks in total). The demand/severity level for each task is within the range of $[0,4]$. At the beginning of the operation, all tasks start with the severity level of 4, and the operation is considered complete when the severity levels of all tasks are 0. Each agent is equipped with two task-specific attributes (firefighting and rescue) and two perception-related attributes (sensing and communication). The joint capability level for each attribute is within the range of $[0,5]$, and it is assumed that joint capability is computed as the addition of the capabilities of the agents assigned to the same task and capped at level 5. Higher task-specific capabilities would reduce the task demand with higher probabilities, and task levels would not increase once they reach level 0. The dependency between tasks is also modelled in task transition probabilities, where the probability for rescuing demand to reduce is 0 when fire level is high ($>2$) no matter how capable the agents are. An agent is only able to observe the task level at the assigned task, making the problem partially observable. Higher sensing and capabilities would help agents perceive the environment more accurately, and the observed information broadcasted by the agents with higher communication capabilities would be trusted more. Here, the sensing and communication capabilities for all agents are set to level 3 (moderately high accuracy) since perception accuracy is not the focus of this study. With a proper designed reward function, agents would reduce the task demands efficiently by cooperating on the same task or simultaneously working on different tasks. Meanwhile, without compromising the operation effectiveness, agents' load is reduced by appropriate idling behaviors and avoiding extra capability usage and unnecessary assignments.

\subsection{Effect of Idle Incentive and Reassignment Penalty on Load Management}
\label{sec:R_LM_effect}
In real operations, constantly engaging or being heavily loaded in task execution would increase the risk exposure and reduce the capability of task-handling for agents, which might cause unexpected deficiencies in operations. Training with the proposed HTLM Dec-POMDP framework and the proper designed reward functions allows agents to make intelligent decisions with their preferences and to be prepared for additional/unexpected task loads. In this study, we investigate the impact of varying levels of idle incentive and reassignment penalty in reward functions on team performance and load reduction. We assume that agents have distinct capability levels in firefighting and rescue as shown in Table~\ref{tb:agent_capa}, capabilities do not change throughout the operation, and agents use the same decision model and reward function. As described in Section~\ref{sec:preference}, for $R_{II}$, agents receive positive reward when idling and negative reward if the capability level is 2-level higher than the believed severity level of the assigned task, reflecting penalty on unnecessary assignment. For the task reassignment penalty $T_{RP}$, agents receive negative reward if being reassigned to a task at a different location. While the reward functions can be adjusted by agents to model their preferences in operations, this study focuses on the change in the reward magnitude given to the team instead of the difference in reward functions among the agents.

\begin{figure*}[!t]
    \centering
    \includegraphics[width=0.85\textwidth]{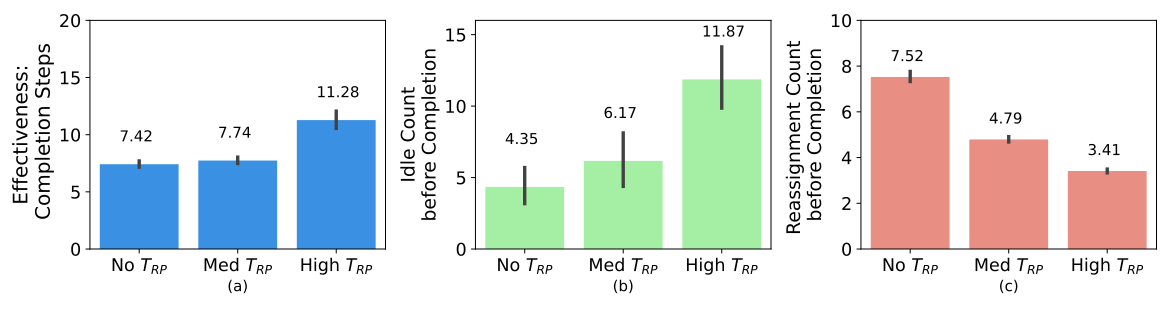}
    \vspace{-3mm}
    \caption{Effect of reassignment penalty $T_{RP}$ on operation effectiveness (a) and load conservation (b), (c). Bar plots show the mean and the standard error for each performance metric evaluated using 500 realizations.}
    \label{fig:reassign_effect}
\end{figure*}

\begin{table}[ht]
\caption{Capabilities for a Heterogeneous Team}
\centering
    \begin{tabular}{|c|c|c|}
         \hline
         Agent ID & Firefighting & Rescue \\
         \hline
         1 & 3 & 1\\
         2 & 3 & 2\\
         3 & 2 & 0\\
         4 & 0 & 1\\
         5 & 1 & 1\\
         \hline
    \end{tabular}
\label{tb:agent_capa}
\end{table}

We train the team of agents with 6 different combinations of idle incentive $R_{II}$ and reassignment penalty $T_{RP}$, and evaluate the trained agents with two performance metrics as shown in Fig.~\ref{fig:idle_effect} with and without $R_{II}$ under (a) no, (b) medium, and (c) high reassignment penalty $T_{RP}$. 
The agents are evaluated in 500 operation trials, and each operation has 30 steps. The first performance metric is operation effectiveness in terms of the steps taken to complete all tasks. When training with no and medium $T_{RP}$, team effectiveness remains at the same level and is not affected by the idling behaviors induced by idle incentive. When $T_{RP}$ is too high, agents are less willing to move and an additional idle incentive further decreases the effectiveness. The second metric is the unused capability per agent per step. Whenever an agent decides to idle, the capability of the agent is considered as unused. On average, each agent has total task-handling capability level 2.8 (1.8 in firefighting and 1 in rescue) as indicated in Table \ref{tb:agent_capa}, and most of the capability $\approx 71\%$ is saved throughout an operation compared to $\approx 2.5\%$ when there is no idle incentives as shown in Fig. \ref{fig:idle_effect}. The amount of unused capability without $R_{II}$ is not always 0 because of the randomness and agent's exploratory behavior during training.

Next, we showcase the impact of reassignment penalty $T_{RP}$ on load management in Fig.~\ref{fig:reassign_effect} by fixing the amplitude of idle incentive. Giving more reassignment penalty during training incentivizes agents to reduce reassignment or overall traversing cost if the penalty is the distance between two task locations. Agents are trained using different levels of reassignment penalty (no, medium and high), and in evaluation, three metrics are investigated: (a) operation effectiveness, (b) team idle count before completion, and (c) team reassignment count before completion (Fig.~\ref{fig:reassign_effect}). For training with medium $T_{RP}$ levels, the team ends up with more idling behaviors and less frequent reassignment before completing all tasks without compromising the operation effectiveness compared to no $T_{RP}$ penalty case (Fig. \ref{fig:reassign_effect}(a)). When $T_{RP}$ is too high during training, agents prefer to stay at the same task or stay idling, but operation effectiveness cannot be guaranteed.

\begin{figure}[!t]
    \begin{subfigure}{0.48\textwidth}
        \centering
        \includegraphics[width=1\textwidth]{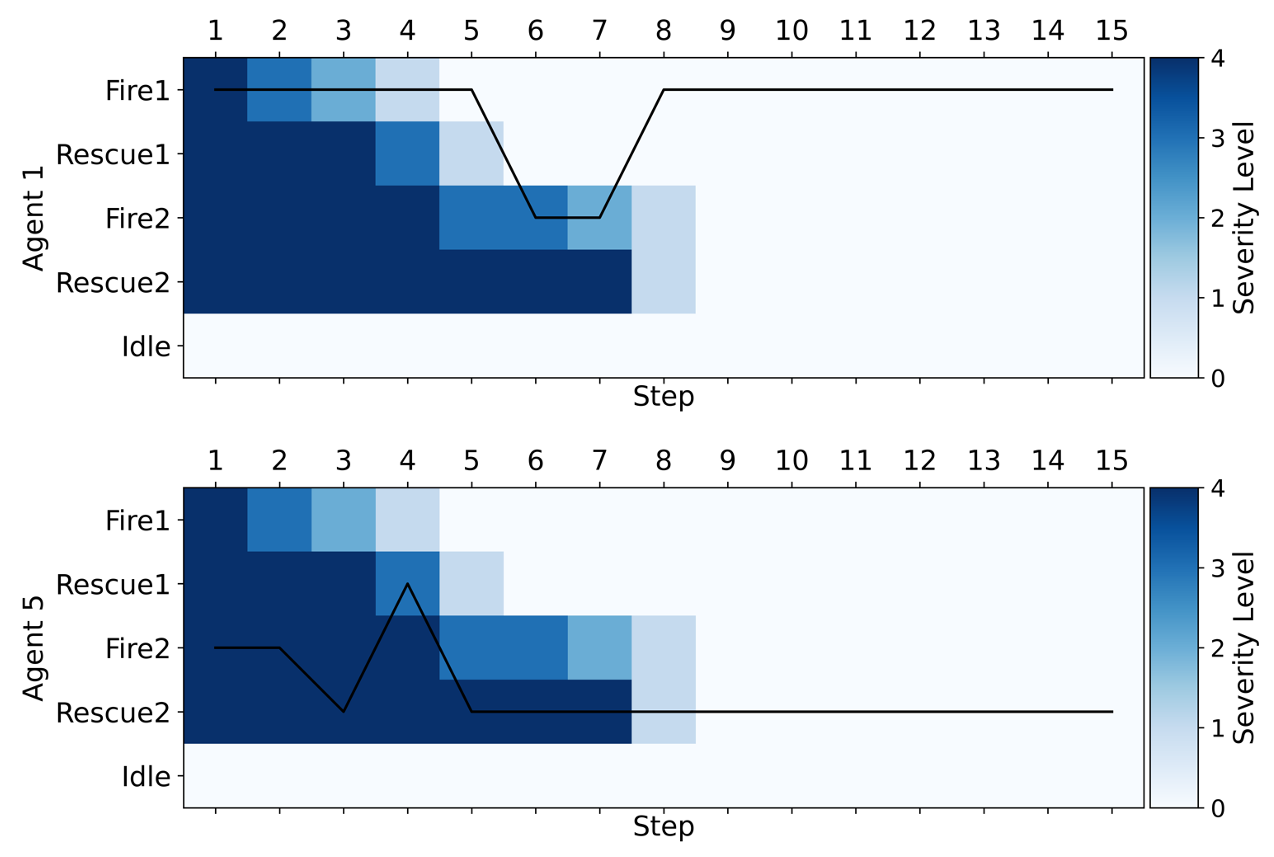}
        \vspace{-6mm}
        \caption{Decisions of Agent 1 and 5 trained with no $R_{II}$ and no $T_{RP}$}
        \label{fig:reg_behavior}
    \end{subfigure}
    \hfill
    \vspace{3mm}
    \begin{subfigure}{0.48\textwidth}
        \centering
        \includegraphics[width=1\textwidth]{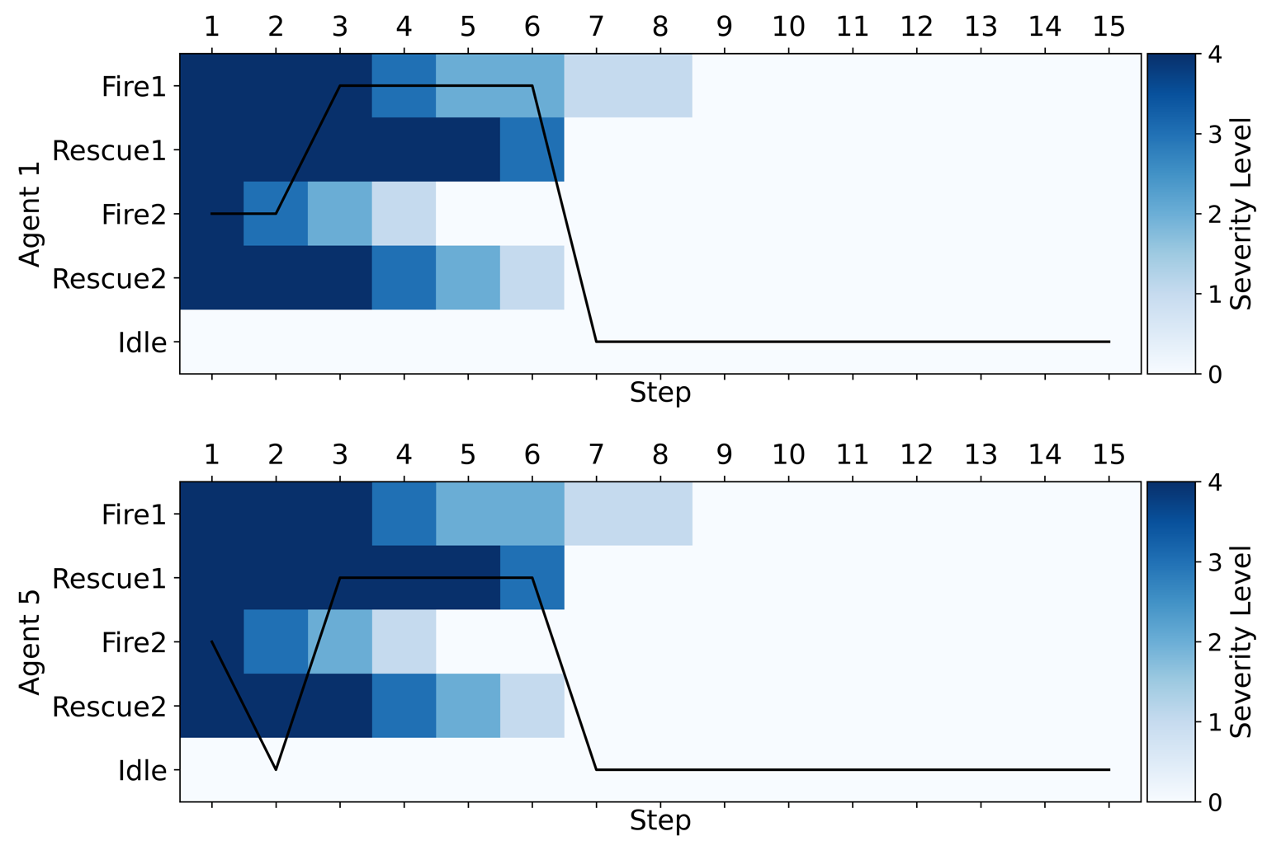}
        \vspace{-6mm}
        \caption{Decisions of Agent 1 and 5 trained with $R_{II}$ and medium $T_{RP}$}
        \label{fig:idle_behavior}
    \end{subfigure}
    \caption{Demonstration of task allocation in an operation, where agents are trained (a) with no $R_{II}$ and no $T_{RP}$, and (b) with $R_{II}$ and medium $T_{RP}$. Black lines indicate agent decisions for the first 15 steps, and color blocks represent the actual severity levels for each task ranging from 0 to 4.}
    \label{fig:agent_behavior}
\end{figure}

To visualize the learned agent behaviors in our proposed HTLM Dec-POMDP, we demonstrate the intelligent task allocation decisions made by two agents in the team shown in Fig.~\ref{fig:idle_behavior} and compare to the decisions trained with no idle incentive or reassignment penalty shown in Fig.~\ref{fig:reg_behavior}. With the help of idle incentive and reassignment penalty, agent task load is reduced by idling when its capability is not needed and when all tasks are completed. Agent 1 with firefighting capability of 3 gets assigned to fire tasks when fire levels are above 1 and idles when fire level is 1 (steps 7 and 8). At steps 2, 7, and 8, Agent 5 with both firefighting and rescue capability of 1 stays idle, letting other agents work on the remaining tasks. Furthermore, the reassignment count for both Agent 1 and 5 is only 1 (step 3 for Agent 1 and Agent 5 in Fig.~\ref{fig:idle_behavior}), while the reassignment count is 2 for both agents trained with no $T_{RP}$ (steps 6, 8 for Agent 1 and step 4, 5 for Agent 5 in Fig.~\ref{fig:idle_behavior}).

\subsection{Agent Importance}
Before training a team of agents for an operation, it is complicated to decide which agents should be chosen as team members. Random team composition based on agent capabilities might cost additional training time if the selected agents are not capable enough to accomplish tasks efficiently or cost too much capability usage if they are all being deployed. The proposed HTLM Dec-POMDP framework could provide insights on team formation and agent importance in terms of their trained behaviors and available capabilities. Identifying the importance of agents could help selecting team members and inferring the team resilience when one of the trained agents is lost or not performing well in the team. A metric for agent importance depending on capability usage and task urgency is proposed in the following equations:

\begin{equation}
    \label{eq:imp}
    \zeta=\sum_{k=1}^K\omega_k u_k,
\end{equation}

\begin{equation}
    \label{eq:capa_usage}
    u_k=\frac{1}{E}\sum_{i=1}^{E}c_k\textbf{1}(\pi^*=a_k|i),
\end{equation}

\begin{equation}
    \label{eq:task_weight}
    \bar{\omega}_k=\sum_{i=1}^{E}\sum_{j=1}^{N}\textbf{1}(\pi_j^*=a_k|i),
\end{equation}

\begin{equation}
    \label{eq:task_weight2}
    \omega_k=\frac{\bar{\omega}_k}{\sum_{k=1}^K \bar{\omega}_k}.
\end{equation}

Here, the importance metric $\zeta:=f(U,W|C,\pi^*,n_s)$ for each agent is defined as a function of its capability usage on each task type $U=[u_1,\dots,u_K]$ depending on its available task-specific capabilities $C=[c_1,\dots,c_K]$ and the task urgency weight $W=[\omega_1,\dots,\omega_K]$, where $K$ is the number of task types in an operation. The importance is estimated through $E$ simulated operation episodes and the trained optimal agent policy $\pi^*$ (i.e. task allocation decisions). The capability usage for each task type $u_k\in U$ is computed as the average capability used in $E$ episodes as shown in Eq. (\ref{eq:capa_usage}), where $a_k$ indicates the agent is allocated to task type $k\in{1,\dots,K}$. However, tasks have different dynamics, and some tasks are harder to resolve than others. Taking our fire and rescue operation as an example, for each step, fire severity can only be reduced by one level at most, and multiple people can be rescued only when the fire level is low, which naturally makes firefighting tasks harder and more urgent. Capability usage only accounts for the resource needed for each task, but an important agent would spend more capability on more urgent tasks. Instead of making arbitrary weights for task urgency, we assume that the task allocated more frequently is more urgent. With a team of $N$ agents and $E$ episodes, the estimated task urgency weight is computed as Eq. (\ref{eq:task_weight}) and normalized using Eq. (\ref{eq:task_weight2}). Then, the importance for each agent is calculated using Eq. (\ref{eq:imp}).

\begin{table}[hb]
\caption{Agent Importance for Three Teams}
\centering
\begin{subtable}[b]{0.48\textwidth}
    \centering
    \begin{tabular}{|c|c|c|c|c|c|}
         \hline
         Agent & $C_{fire}$ & $C_{rescue}$ & $U_{fire}$ & $U_{rescue}$ & $\zeta$\\
         \hline
         1 & 3 & 1 & 21.19 & 0.04 & 13.90\\
         2 & 3 & 2 & 16.92 & 3.18 & 12.18\\
         3 & 2 & 0 & 13.26 & 0 & 8.69\\
         \textbf{4} & 0 & 1 & 0 & 7.17 & \textbf{2.47}\\
         5 & 1 & 1 & 3.64 & 3.32 & 3.57\\
         \hline
    \end{tabular}
    \vspace{-2mm}
  \caption{Heterogeneous Team $\omega_{fire}=0.66, \omega_{rescue}=0.34$}
  \label{tb:heter_imp}
\end{subtable}
  \hfill
\vspace{2mm}
\begin{subtable}[b]{0.48\textwidth}
    \centering
    \begin{tabular}{|c|c|c|c|c|c|}
         \hline
         Agent & $C_{fire}$ & $C_{rescue}$ & $U_{fire}$ & $U_{rescue}$ & $\zeta$\\
         \hline
         \textbf{1} & 3 & 1 & 3.44 & 0 & \textbf{2.58}\\
         2 & 3 & 1 & 22.38 & 0 & 16.75\\
         3 & 1 & 1 & 10.80 & 0 & 8.08\\
         4 & 2 & 2 & 3.70 & 4.26 & 3.84\\
         5 & 2 & 2 & 8.99 & 9.42 & 9.10\\
         \hline
    \end{tabular}
    \vspace{-2mm}
  \caption{Semi-Heterogeneous Team $\omega_{fire}=0.75, \omega_{rescue}=0.25$}
  \label{tb:semiheter_imp}
\end{subtable}
    \hfill
\vspace{2mm}
\begin{subtable}[b]{0.48\textwidth}
    \centering
    \begin{tabular}{|c|c|c|c|c|c|}
         \hline
         Agent & $C_{fire}$ & $C_{rescue}$ & $U_{fire}$ & $U_{rescue}$ & $\zeta$\\
         \hline
         1 & 2 & 1 & 17.91 & 1.49 & 12.49\\
         2 & 2 & 1 & 12.68 & 4.05 & 9.84\\
         \textbf{3} & 2 & 1 & 2.52 & 3.20 & \textbf{2.74}\\
         4 & 2 & 1 & 12.17 & 3.00 & 9.15\\
         5 & 2 & 1 & 14.60 & 2.99 & 10.77\\
         \hline
    \end{tabular}
    \vspace{-2mm}
  \caption{Homogeneous Team $\omega_{fire}=0.67, \omega_{rescue}=0.33$}
  \label{tb:homo_imp}
\end{subtable}
\label{tb:agent_imp}
\end{table}

In this analysis, we use the same operation scenario with $h=30$ steps, consisting of two task types (fire and rescue) at two locations and five agents. The importance score is ranging from 0 if the agent always idles to $c_kh$ if all agents always work on the same task type. In addition to a team of five agents with heterogeneous capabilities as described in Table \ref{tb:agent_capa}, a semi-heterogeneous team and a homogeneous team are considered. The available capabilities $(C_{fire},C_{rescue})$, used capabilities $(U_{fire},U_{rescue})$, and the importance $\zeta$ of each agent for three teams are summarized in Table \ref{tb:agent_imp}. The agent with the least importance score for each team is highlighted in bold.

\begin{figure}[!t]
    \centering
    \begin{subfigure}{0.48\textwidth}
        \centering
        \includegraphics[width=1\textwidth]{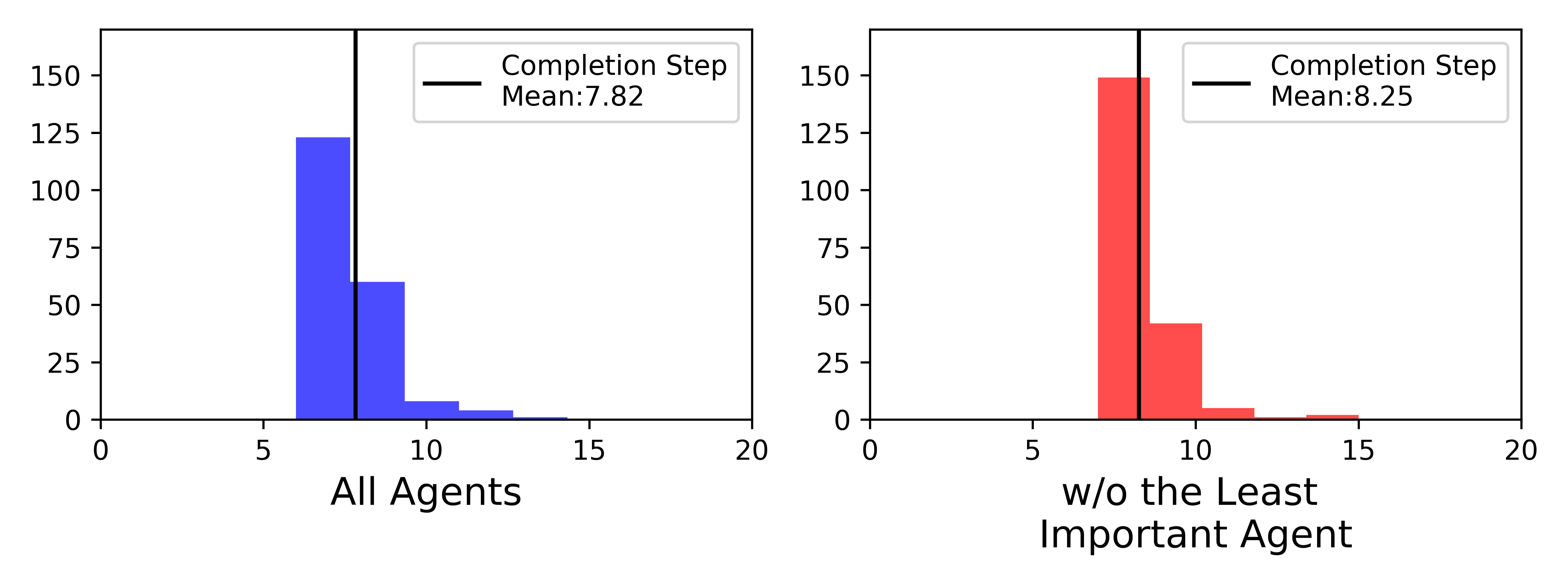}
        \label{fig:semi_imp}
    \end{subfigure}
    \hfill
    \vspace{-4mm}
    \begin{subfigure}{0.48\textwidth}
        \centering
        \caption{Heterogeneous Team}
        \includegraphics[width=1\textwidth]{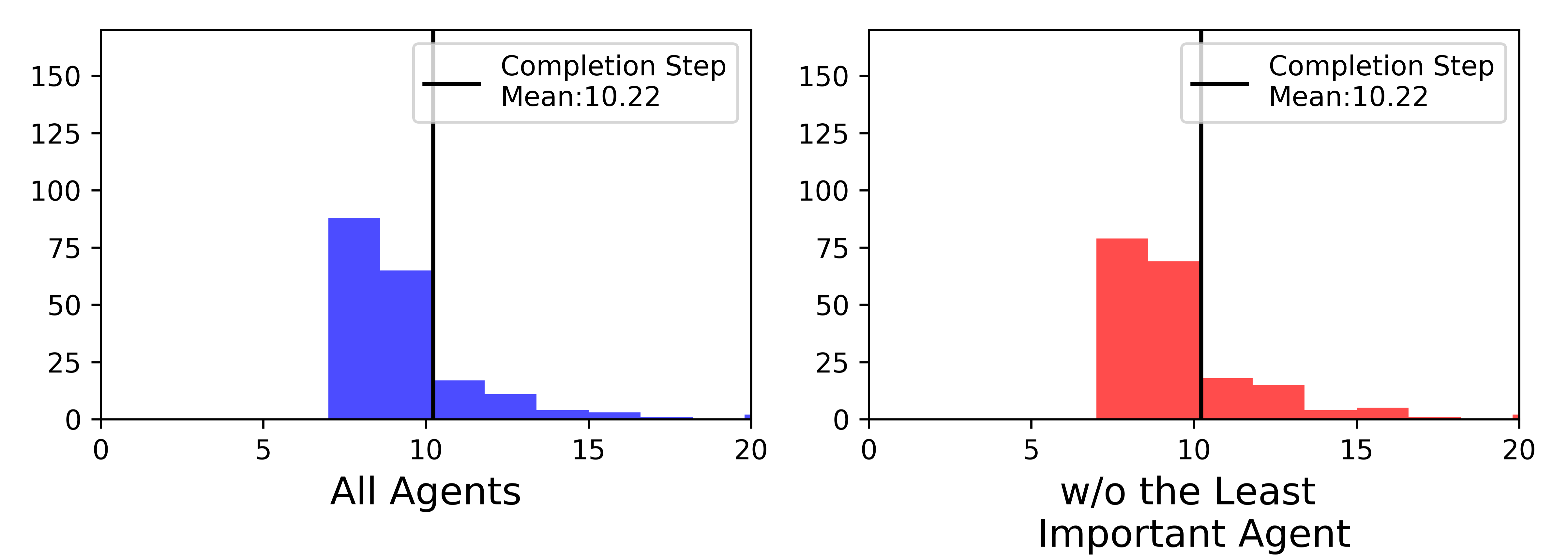}
        \label{fig:semi_imp}
    \end{subfigure}
    \hfill
    \vspace{-4mm}
    \begin{subfigure}{0.48\textwidth}
        \centering
        \caption{Semi-Heterogeneous Team}
        \includegraphics[width=1\textwidth]{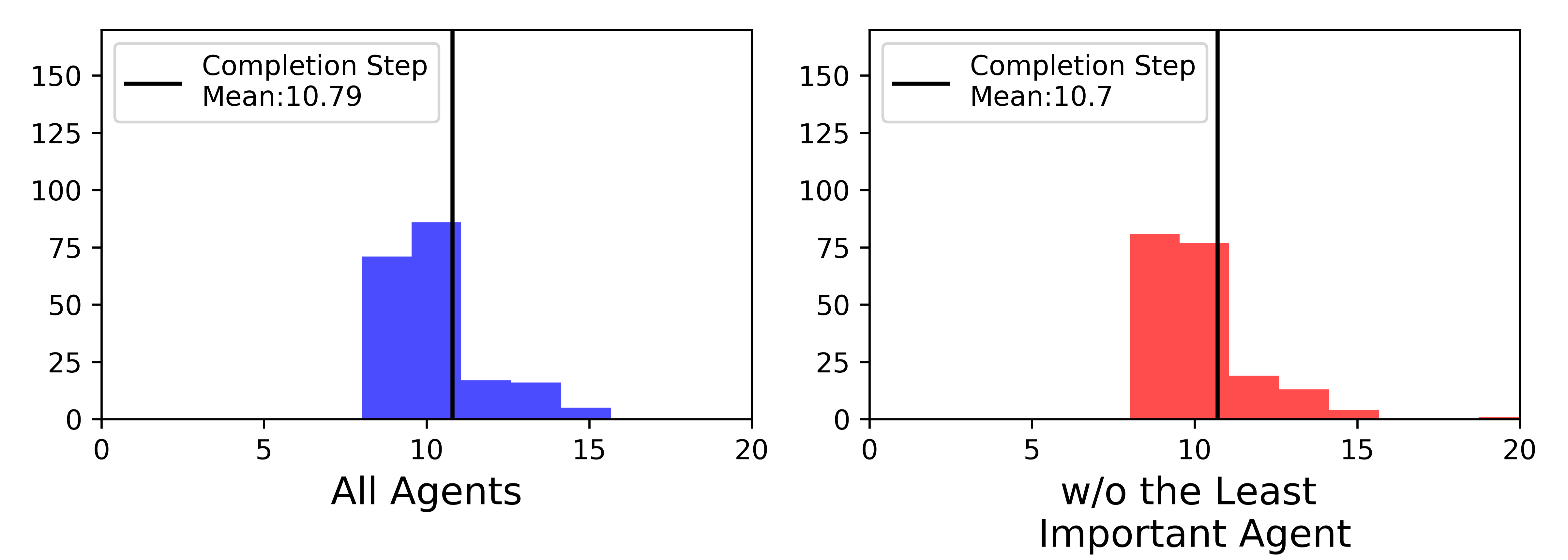}
        \label{fig:homo_imp}
    \end{subfigure}
    \hfill
    \vspace{-4mm}
    \begin{subfigure}{0.48\textwidth}
        \centering
        \caption{Homogeneous Team}
    \end{subfigure}
    \vspace{-5mm}
    \caption{Operation effectiveness comparison between having all agents (blue) and inactivating the least important agent (red) for (a) Heterogeneous Team, (b) Semi-Heterogeneous Team, (c) Homogeneous Team. Each histogram shows 200 realizations with the mean completion step indicated.}
    \label{fig:imp_compare}
\end{figure}

To validate the agent importance, without re-training the agents, we set one of the agents in a team to always idle and evaluate the operation effectiveness described as completion step compared to having all agents. Fig. \ref{fig:imp_compare} suggests that the operation effectiveness is not affected after inactivating the the least important agent indicated in Table \ref{tb:agent_imp} for all three teams. Inactivating the most important agent for all teams would lead to operation failure (i.e. unable to accomplish tasks within 30 steps). It is worth noting that the agent importance metric is not simply the indicator of its capability but depends also on the learned decision process during the training in interaction with other agents. Consequently, agents with the same capability might have different importance (Table \ref{tb:homo_imp}) and agents with higher capabilities might be the least important and unnecessary to execute any tasks (Table \ref{tb:semiheter_imp}). Hence, the agent importance is the metric for the significance of the agent in a learned team collaboration. We have demonstrated that the importance metric for agents trained with the HTLM Dec-POMDP framework could help to infer which agent is more significant in collaboration, how much capability could be saved without affecting the operation effectiveness, and consequently how much additional demand can be handled by the team by fully activating all agents. 

Although this study focuses on fire suppression and rescue operations, the formulation is general enough to model different types of task dynamics  and various team compositions ranging from homogeneous to heterogeneous teams. The formulation could be generalized to autonomy applications in different fields including but not limited to warehouse management and task scheduling in manufacturing.


%% file: sections/Conclusions.tex
In this work, we formulate a decentralized intelligent decision-making framework for multi-agent teams to learn efficient task allocation strategies with load management through deep reinforcement learning. The proposed framework (HTLM Dec-POMDP) models idling as an option of allocation decision and encourages idling behaviors to conserve task load and available resources by customizing individual reward functions based on agent preferences. Training multi-agent teams with the HTLM Dec-POMDP framework allows agents to reduce task load without compromising the operation effectiveness and provides the opportunity to investigate team member selection, team resilience, and precaution preparation. Results show that properly defined reward functions can achieve load reduction and resource conservation while maintaining team performance. A metric for agent importance is defined to measure agent significance within the team and to suggest team formation and reliability.

The proposed method serves as an incentive for developing intelligent multi-agent task allocation systems with the consideration of load management, which could be beneficial for designing management-level strategies in complex operations for human-autonomy teams. The method could potentially suggest decisions to human agents in real practices without being overloaded. 
In large-scale problems, as more tasks and agents are involved, a logistic of subteam formation and a hierarchical task planning method are desired to enhance computational efficiency. Moreover, the study of agent importance demonstrates asymmetric behaviors in collaboration with respect to capabilities, which potentially opens the discussion on studying collaborative teaming behaviors in the future.

%% file: main.bbl
\begin{thebibliography}{10}
\providecommand{\url}[1]{#1}
\csname url@samestyle\endcsname
\providecommand{\newblock}{\relax}
\providecommand{\bibinfo}[2]{#2}
\providecommand{\BIBentrySTDinterwordspacing}{\spaceskip=0pt\relax}
\providecommand{\BIBentryALTinterwordstretchfactor}{4}
\providecommand{\BIBentryALTinterwordspacing}{\spaceskip=\fontdimen2\font plus
\BIBentryALTinterwordstretchfactor\fontdimen3\font minus
  \fontdimen4\font\relax}
\providecommand{\BIBforeignlanguage}[2]{{%
\expandafter\ifx\csname l@#1\endcsname\relax
\typeout{** WARNING: IEEEtran.bst: No hyphenation pattern has been}%
\typeout{** loaded for the language `#1'. Using the pattern for}%
\typeout{** the default language instead.}%
\else
\language=\csname l@#1\endcsname
\fi
#2}}
\providecommand{\BIBdecl}{\relax}
\BIBdecl

\bibitem{MATAreview}
S.~Saravanan, K.~C. Ramanathan, R.~MM, and M.~N. Janardhanan, ``Review on
  state-of-the-art dynamic task allocation strategies for multiple-robot
  systems,'' \emph{Industrial Robot}, vol.~47, pp. 929--942, 2020.

\bibitem{MATA2}
K.~Lerman, C.~Jones, A.~Galstyan, and M.~J. Matarić, ``Analysis of dynamic
  task allocation in multi-robot systems,'' \emph{The International Journal of
  Robotics Research}, vol.~25, no.~3, pp. 225--241, 2006.

\bibitem{MATA3}
M.~Gini, ``Multi-robot allocation of tasks with temporal and ordering
  constraints,'' \emph{Proceedings of the AAAI Conference on Artificial
  Intelligence}, vol.~31, no.~1, 2017.

\bibitem{DEC-POMDP}
F.~A. Oliehoek, M.~T.~J. Spaan, and N.~A. Vlassis, ``Optimal and approximate
  q-value functions for decentralized pomdps,'' \emph{Journal Of Artificial
  Intelligence Research}, vol.~32, pp. 289--353, 2011.

\bibitem{HOOSHANGI2017}
N.~Hooshangi and A.~{Asghar Alesheikh}, ``Agent-based task allocation under
  uncertainties in disaster environments: An approach to interval
  uncertainty,'' \emph{International Journal of Disaster Risk Reduction},
  vol.~24, pp. 160--171, 2017.

\bibitem{DQN}
V.~Mnih, K.~Kavukcuoglu, D.~Silver, A.~Graves, I.~Antonoglou, D.~Wierstra, and
  M.~Riedmiller, ``Playing atari with deep reinforcement learning,''
  \emph{CoRR}, vol. abs/1312.5602, pp. 1--9, 2013.

\bibitem{omidshafiei2017deep}
S.~Omidshafiei, J.~Pazis, C.~Amato, J.~How, and J.~Vian, ``Deep decentralized
  multi-task multi-agent reinforcement learning under partial observability,''
  \emph{Proceedings of the 34th International Conference on Machine Learning},
  vol.~70, pp. 2681--2690, 2017.

\bibitem{HT-DEC-POMDP}
H.~Wu, A.~Ghadami, A.~E. Bayrak, J.~M. Smereka, and B.~I. Epureanu, ``Impact of
  heterogeneity and risk aversion on task allocation in multi-agent teams,''
  \emph{IEEE Robotics and Automation Letters}, vol.~6, no.~4, pp. 7065--7072,
  2021.

\bibitem{Nair2013}
R.~Nair, M.~Tambe, M.~Yokoo, P.~D., and S.~Marsella, ``Taming decentralized
  pomdps: towards efficient policy computation for multiagent settings,'' 2013,
  pp. 705--711.

\bibitem{emam2020adaptive}
Y.~Emam, S.Mayya, G.~Notomista, A.~Bohannon, and M.~Egerstedt, ``Adaptive task
  allocation for heterogeneous multi-robot teams with evolving and unknown
  robot capabilities,'' \emph{International Conference on Robotics and
  Automation}, pp. 7719--7725, 2020.

\bibitem{QMDPNet}
P.~Karkus, D.~Hsu, and W.~Lee, ``Qmdp-net: Deep learning for planning under
  partial observability,'' \emph{International Conference on Neural Information
  Processing Systems}, p. 4697–4707, 2017.

\bibitem{MATARL}
L.~Bu{\c{s}}oniu, R.~Babu{\v{s}}ka, and B.~De~Schutter, \emph{Multi-agent
  Reinforcement Learning: An Overview}.\hskip 1em plus 0.5em minus 0.4em\relax
  Berlin, Heidelberg: Springer Berlin Heidelberg, 2010, pp. 183--221.

\bibitem{ResourceManage}
H.~Mao, M.~Alizadeh, I.~Menache, and S.~Kandula, ``Resource management with
  deep reinforcement learning,'' \emph{Proceedings of the 15th ACM Workshop on
  Hot Topics in Networks}, p. 50–56, 2016.

\bibitem{ResourceAllocation}
S.~Banerjee and J.~Hecker, ``A multi-agent system approach to load-balancing
  and resource allocation for distributed computing,'' \emph{Complex Systems
  Digital Campus 2015 World eConference on Complex Systems}, pp. 41--54, 12
  2017.

\bibitem{ResAlloMA}
N.~Creech, N.~C. Pacheco, and S.~Miles, ``Resource allocation in dynamic
  multiagent systems,'' \emph{CoRR}, vol. abs/2102.08317, pp. 1--22, 2021.

\bibitem{MAWorkload}
M.~Dadvar, S.~Moazami, H.~R. Myler, and H.~Zargarzadeh, ``Multi-agent task
  allocation in complementary teams: {A} hunter and gatherer approach,''
  \emph{CoRR}, vol. abs/1912.05748, pp. 1--15, 2019.

\bibitem{AdaptiveWLA}
T.~Mina, S.~S. Kannan, W.~Jo, and B.~Min, ``Adaptive workload allocation for
  multi-human multi-robot teams for independent and homogeneous tasks,''
  \emph{IEEE Access}, vol.~8, pp. 152\,697--152\,712, 2020.

\bibitem{AuctionTAforMR}
E.~Schneider, E.~Sklar, S.~Parsons, and A.~Ozgelen, ``Auction-based task
  allocation for multi-robot teams in dynamic environments,'' \emph{Towards
  Autonomous Robotic Systems}, pp. 246--257, 2015.

\bibitem{Noureddine2017MultiagentDR}
D.~B. Noureddine, A.~Gharbi, and S.~Ahmed, ``Multi-agent deep reinforcement
  learning for task allocation in dynamic environment,'' \emph{International
  Conference on Software Technologies}, 2017.

\bibitem{idleinoccu}
A.~Maoudj, B.~Bouzouia, A.~Hentout, and R.~Toumi, ``Multi-agent approach for
  task allocation and scheduling in cooperative heterogeneous multi-robot team:
  Simulation results,'' \emph{IEEE 13th International Conference on Industrial
  Informatics}, pp. 179--184, 2015.

\bibitem{prodSche}
F.~Shrouf, J.~Ordieres-Meré, A.~García-Sánchez, and M.~Ortega-Mier,
  ``Optimizing the production scheduling of a single machine to minimize total
  energy consumption costs,'' \emph{Journal of Cleaner Production}, vol.~67,
  pp. 197--207, 2014.

\bibitem{energyManage}
V.-H. Bui, A.~Hussain, and H.-M. Kim, ``A multiagent-based hierarchical energy
  management strategy for multi-microgrids considering adjustable power and
  demand response,'' \emph{IEEE Transactions on Smart Grid}, vol.~9, no.~2, pp.
  1323--1333, 2018.

\bibitem{cloudComp}
M.~Gawali and S.~Shinde, ``Task scheduling and resource allocation in cloud
  computing using a heuristic approach,'' \emph{Journal of Cloud Computing},
  vol.~7, no.~4, 02 2018.

\bibitem{resilientTA}
S.~Mayya, D.~S. D’antonio, D.~Saldaña, and V.~Kumar, ``Resilient task
  allocation in heterogeneous multi-robot systems,'' \emph{IEEE Robotics and
  Automation Letters}, vol.~6, no.~2, pp. 1327--1334, 2021.

\end{thebibliography}
